\documentclass[twocolumn,aps,prl,superscriptaddress,showpacs,secnumroman,showkeys]{revtex4}
\usepackage{amsmath,bm}%
\usepackage{graphicx}%
\usepackage{epsf,epsfig,latexsym}
\usepackage{color}
\begin{document}
\title{The Nuclear Matter Symmetry Energy  at $0.03\leq \rho/\rho_0\leq 0.2$}  

\author{R. Wada}
\affiliation{Institute of Modern Physics HIRFL, Chinese Academy of Sciences, 
Lanzhou, 730000, China}
\affiliation{Cyclotron Institute, Texas A$\&$M University, College Station, 
Texas 77843}
\author{K. Hagel}
\affiliation{Cyclotron Institute, Texas A$\&$M University, College Station, 
Texas 77843}
\author{L. Qin}
\affiliation{Cyclotron Institute, Texas A$\&$M University, College Station, 
Texas 77843}
\author{J. B. Natowitz}
\affiliation{Cyclotron Institute, Texas A$\&$M University, College Station, 
Texas 77843}
\author{Y. G. Ma}
\affiliation{Shanghai Institute of Nuclear Research,
Chinese Academy of Sciences, Shanghai 201800, China}
\author{G. R\"opke}
\affiliation{University of Rostock, FB Physik, Rostock, Germany }
\author{S. Shlomo}
\affiliation{Cyclotron Institute, Texas A$\&$M University, College Station, 
Texas 77843}
\author{A. Bonasera}
\affiliation{Cyclotron Institute, Texas A$\&$M University, College Station, 
Texas 77843}
\affiliation{Laboratori Nazionali del Sud, INFN, via Santa Sofia, 62, 95123 
Catania, Italy}
\author{S. Typel}
\affiliation{GSI Helmholtzzentrum f¨ur Schwerionenforschung GmbH,Theorie, 
Planckstra\ss e 1, D-64291 Darmstadt, Germany}
\author{Z. Chen}
\affiliation{Cyclotron Institute, Texas A$\&$M University, College Station, 
Texas 77843}
\affiliation{Institute of Modern Physics HIRFL, Chinese Academy of Sciences, 
Lanzhou, 730000, China}
\author{M. Huang}
\affiliation{Cyclotron Institute, Texas A$\&$M University, College Station, 
Texas 77843}
\affiliation{Institute of Modern Physics HIRFL, Chinese Academy of Sciences, 
Lanzhou, 730000, China}
\author{J. Wang}
\affiliation{Cyclotron Institute, Texas A$\&$M University, College Station, 
Texas 77843}
\affiliation{Institute of Modern Physics HIRFL, Chinese Academy of Sciences, 
Lanzhou, 730000, China}
\author{H. Zheng}
\affiliation{Cyclotron Institute, Texas A$\&$M University, College Station, 
Texas 77843}
\author{S. Kowalski}
\affiliation{Institute of Physics, Silesia University, Katowice, Poland.}
\author{C. Bottosso}
\affiliation{Cyclotron Institute, Texas A$\&$M University, College Station, 
Texas 77843}
\author{M. Barbui}
\affiliation{Cyclotron Institute, Texas A$\&$M University, College Station, 
Texas 77843}
\author{M. R. D. Rodrigues}
\affiliation{Cyclotron Institute, Texas A$\&$M University, College Station, 
Texas 77843}
\author{K. Schmidt}
\affiliation{Cyclotron Institute, Texas A$\&$M University, College Station, 
Texas 77843}
\author{D. Fabris}
\affiliation{Dipartamento di Fisica dell~Universita di Padova and INFN Sezione di Padova, Padova, Italy}
\author{M. Lunardon}
\affiliation{Dipartamento di Fisica dell~Universita di Padova and INFN Sezione di Padova, Padova, Italy}
\author{S. Moretto}
\affiliation{Dipartamento di Fisica dell~Universita di Padova and INFN Sezione di Padova, Padova, Italy}
\author{G. Nebbia}
\affiliation{Dipartamento di Fisica dell~Universita di Padova and INFN Sezione di Padova, Padova, Italy}
\author{S. Pesente}
\affiliation{Dipartamento di Fisica dell~Universita di Padova and INFN Sezione di Padova, Padova, Italy}
\author{V. Rizzi}
\affiliation{Dipartamento di Fisica dell~Universita di Padova and INFN Sezione di Padova, Padova, Italy}
\author{G. Viesti}
\affiliation{Dipartamento di Fisica dell~Universita di Padova and INFN Sezione di Padova, Padova, Italy}
\author{M. Cinausero}
\affiliation{INFN Laboratori Nazionali di Legnaro, Legnaro, Italy}
\author{G. Prete}
\affiliation{INFN Laboratori Nazionali di Legnaro, Legnaro, Italy}
\author{T. Keutgen}
\affiliation{FNRS and IPN, Universit\'e Catholique de Louvain, B-1348 
Louvain-Neuve, Belgium}
\author{Y. El Masri}
\affiliation{FNRS and IPN, Universit\'e Catholique de Louvain, B-1348 
Louvain-Neuve, Belgium}
\author{Z. Majka}
\affiliation{Smoluchowski Institute of Physics, Jagiellonian University, 
Krakow, Poland}

\date{\today}

\begin{abstract}
Measurements of the density dependence of the Free symmetry energy 
in low density clustered matter have been extended using the NIMROD 
multi-detector at Texas A\&M University. Thermal coalescence models 
were employed to extract densities, $\rho$, and temperatures, $T$, 
for evolving systems formed in collisions of 47 $A$ MeV 
$^{40}$Ar + $^{112}$Sn ,$^{124}$Sn  and $^{64}$Zn + $^{112}$Sn , 
$^{124}$Sn.  Densities of  $0.03 \leq \rho/\rho_0 \leq 0.2$ and 
temperatures in the range 5 to 10 MeV have been sampled. The Free symmetry 
energy coefficients are found to be in good agreement with values 
calculated using a quantum statistical model. Values of the 
corresponding symmetry energy coefficient are derived from the 
data using entropies derived from the model.  
\end{abstract}

\pacs{25.70.Pq}

\keywords{Intermediate Heavy ion reactions, chemical equilibrium, neutron and proton chemical potential, quantum statisitcal model calculations}

\maketitle
 
\section{I. Introduction}
The symmetry energy of nuclear matter is a fundamental ingredient in the 
investigation of nuclear and astrophysical phenomena. 
The symmetry energy characterizes the dependence of the nuclear binding 
energy on the asymmetry $\delta = (N-Z)/A$ where  $Z$ and $N$ are the 
proton and neutron numbers, and $A = N+Z$.  As a general representation 
of the symmetry energy coefficient we use the definition
\begin{equation}
E_{\rm sym}(\rho,T) = \frac{1}{2}\big(E(\rho,1,T) + E(\rho,-1,T)\big) - E(\rho,0,T)
\end{equation}
where $E(\rho,\delta,T)$ is the energy per nucleon of nuclear matter with 
density $\rho$, asymmetry $\delta$, and temperature $T$.  If a quadratic 
dependence is assumed this  definition becomes identical to the frequently 
used alternative of taking the second derivative of $E(\rho,\delta,T)$ with 
respect to the asymmetry $\delta=0$.  Our empirical knowledge of the symmetry 
energy near the saturation density, $\rho_0$, is based primarily on the 
binding energies of nuclei. The Bethe-Weizs\"acker mass formula leads to 
values of about $E_{\rm sym}(\rho_0,0) = 28-34$ MeV for the symmetry energy at 
zero temperature and saturation density $\rho_0 = 0.16$ fm$^{-3}$, if 
surface asymmetry effects are properly taken into 
account~\cite{danielewicz03}.  In contrast to the value 
of $E_{\rm sym}(\rho_0,0)$, the variation of the symmetry energy with density 
and temperature is intensely debated.  Many experimental and theoretical 
investigations have been performed to estimate the behavior of the symmetry 
energy as a function of $\rho$ and $T$.  Recent reviews are given by 
Li \textit{et al.}~\cite{lireview} and by Lattimer and 
Lim ~\cite{lattimer12}.  

In the Fermi energy domain symmetry energy effects have been 
investigated using judiciously chosen observables from nuclear 
reactions~\cite{lireview,lattimer12,baran05,tsang09,shetty07,sfienti09}.
In the theoretical investigations quasiparticle approaches such as the 
Skyrme Hartree-Fock and relativistic mean field (RMF) models or 
Dirac-Brueckner Hartree-Fock (DBHF) calculations are 
used~\cite{lireview,ref2,ref3}.  In such calculations the symmetry energy 
tends to zero in the low-density limit for uniform matter.  However, in 
accordance with the mass action law, cluster formation dominates the structure 
of low-density symmetric matter at low temperatures. Thus at low density the 
symmetry energy changes mainly because additional binding is gained in 
symmetric matter due to formation of clusters and pasta 
structures~\cite{watanabe09}.  As a result the symmetry energy in this 
low-temperature limit has to be equal to the binding energy per nucleon 
associated with the strong interaction of the most bound nuclear cluster.  
A single-nucleon quasiparticle approach cannot account for such structures.  
The correct low-density limit can be recovered only if the formation of 
clusters is properly taken into account, as has previously been shown 
in~\cite{horowitz06} in the context of a virial expansion valid at very low 
densities and in Ref.~\cite{typel10}.  Since, at low density, the symmetry 
energy changes mainly because additional binding is gained in the formation 
of clusters~\cite{kowalski07,natowitz10,roepke09,roepke11,horowitz06}, 
measurements of nucleon and light cluster emission from the participant 
matter which is produced in near Fermi energy heavy ion collisions can be 
employed to probe the EOS at low density and moderate temperatures where 
clustering is important~\cite{kowalski07,natowitz10}.  Our previous data 
demonstrated a large degree of alpha clustering for densities at and below 
$\sim0.05$ times normal nuclear density, $\rho_0$ (0.16 nucleons/fm$^3$) and 
temperatures of 4 to 10 MeV.  Using these data we derived symmetry energy 
coefficients in low density nuclear matter~\cite{kowalski07,natowitz10}.  
The analysis employed the isoscaling technique which compares yields for two 
systems with similar temperatures but different $N/Z$ ratios to determine the 
differences in chemical potentials and symmetry energy~\cite{tsang01,souza08}.
The NIMROD 4$\pi$ multi-detector at Texas A \& M University has now been used 
to extend our measurements to higher densities. Cluster production in 
collisions of 47 $A$ MeV $^{40}$Ar with $^{112,124}$Sn and $^{64}$Zn 
with $^{112,124}$Sn was studied.  We report here determinations of symmetry 
energy coefficients at $0.03 \leq  \rho/\rho_{0}\leq 0.2$ and moderate 
temperatures.  Our results for this expanded range of densities 
are in reasonable agreement with those of a quantum statistical model 
calculation incorporating medium modifications of the cluster binding 
energies \cite{typel10,sumiyoshi95,ref13}. 

\section{Experimental Techniques}
The experiments were performed using beams from the K500 Superconducting 
Cyclotron at the Texas A\&M University Cyclotron Institute, incident on 
targets in the NIMROD detector.  NIMROD consists of a 166 segment charged 
particle array set inside a neutron ball \cite{wuenschel09}.  The charged 
particle array is arranged in 12 rings of Si-CsI telescopes or single CsI 
detectors concentric around the beam axis. The CsI detectors are 1-10 cm 
thick Tl doped crystals read by photomultiplier tubes. A pulse shape 
discrimination method is employed to identify light particles in the 
CsI detectors. For this experiment each of the forward rings included 
two segments having two Si detectors (150 and 500 $\mu$m thick) in front 
of the CsI detectors (super telescopes) and three having one Si 
detector (300 $\mu$m thick). Each super telescope was further divided 
into two sections. Neutron multiplicity was measured with the 4$\pi$ 
neutron detector surrounding the charged particle array. This detector 
is a neutron calorimeter filled with gadolinium doped pseudocumene.  
Thermalization and capture of emitted neutrons in the ball leads to 
scintillation light which is observed with phototubes providing event 
by event determinations of neutron multiplicity.  Further details on 
the detection system, energy calibrations and neutron ball efficiency 
may be found in references \cite{qin08,hagel00}. The combined neutron 
and charged particle multiplicities were employed to select the most 
violent events for subsequent analysis. 

\section{Analysis}
The dynamics of the collision process allow us to probe the nature of the 
intermediate velocity ``nucleon-nucleon'' emission 
source \cite{kowalski07,natowitz10,qin08,hagel00}.  Measurement of emission 
cross sections of nucleons and light clusters together with suitable 
application of a coalescence ansatz \cite{mekjian78} provides the means to 
probe the properties and evolution of the interaction region.  The 
techniques used in the analysis have been detailed in several previous 
publications \cite{kowalski07,natowitz10,qin08,hagel00}  and are described 
briefly below. A notable difference from Refs. \cite{kowalski07} 
and \cite{natowitz10} is the method of density extraction. This is discussed 
more extensively in the following. We emphasize that the event selection is 
on the more violent collisions. Cross section weighting favors mid-range 
impact parameters.

An initial estimation of emission multiplicities at each stage of the 
reaction was made by fitting the observed light particle spectra assuming 
contributions from three sources, a projectile-like fragment (PLF) source, 
an intermediate velocity (IV) source, and a target-like fragment (TLF) source. 
A reasonable reproduction of the observed spectra is achieved. Except for the 
most forward detector rings the data are dominated by particles associated 
with the IV and TLF sources. The IV source velocities are very close to 50\% 
of the beam velocity as seen in many other studies \cite{qin08,hagel00}. 
The observed spectral slopes reflect the evolution dynamics of the source, 
not its internal temperature \cite{bauer95,zheng11}.  For further analysis, 
this IV source is most easily sampled at the intermediate angles where 
contributions from the other sources are minimized.  For the analysis of 
the evolution of the source we have selected the data in ring 9 of the 
NIMROD detector. This ring covered an angular range in the laboratory 
of 38$^\circ$ to 52$^\circ$. An inspection of invariant velocity plots constructed 
for each ejectile and each system, as well as of the results of the 
three-source fit analyses, indicates that this selection of angular 
range minimizes contributions\textbf{ }from secondary evaporative decay 
of projectile like or target like sources. We treat the IV source as a 
nascent fireball created in the participant interaction zone. The expansion 
and cooling of this zone leads to a correlated evolution of density and 
temperature which we probe using particle and cluster observables, yield, 
energy and angle.

\subsection{Temperature}
As in some previous work \cite{kowalski07,natowitz10} we have employed 
double isotope yield ratios \cite{albergo85,kolomiets97} to characterize 
the temperature at a particular emission time. For particles emitted from 
a single source of temperature, $T$, and having a volume Maxwellian spectrum 
the HHe double isotope yield ratio evaluated for particles of equal 
$v_{\rm surf}$ is $\sqrt{9/8}$ times the ratio derived from 
either the integrated particle yields or the yields at a given energy 
above the barrier,  
\begin{equation}
\label{TAlb}
T_{\rm HHe} = \frac{14.3} {\ln{(\sqrt{9/8} (1.59\,\,R_{v_{\rm surf}}))}}.
\end{equation}
If $Y$ represents a cluster yield, $R_{v_{\rm surf}} =Y(^2{\rm H})Y(^4{\rm He})/Y(^3{\rm H})Y(^3{\rm He})$ 
for clusters with the same surface velocity.  The constants 14.3 and 1.59 
reflect binding energy, spin, masses and mass differences of the ejectiles. 
Eq.~(\ref{TAlb}) differs from the usual formulation only by the factor of 
$\sqrt{9/8}$ appearing in the logarithm term in the denominator.

Model studies \cite{shlomo09}  comparing Albergo model \cite{albergo85} temperatures and densities to the 
known input values have shown the double isotope ratio temperatures to be 
relatively robust in this density range.  However the 
densities extracted using the Albergo relationships are useful only at the 
very lowest densities.  Consequently, in this study we have 
employed a different means of density extraction -- the thermal coalescence 
model of Mekjian \cite{mekjian78}. 

\subsection{Density}
To determine the coalescence parameter $P_{0}$, the radius in momentum space, 
from our data we have followed the Coulomb corrected coalescence model formalism 
of Awes \textit{et al.} \cite{awes81} and previously employed by us in 
Ref. \cite{hagel00}. In the laboratory frame the derived 
relationship between the observed cluster and proton differential cross 
sections is   

\begin{eqnarray}
\frac{d^2N(Z,N,E_A)}{dE_A\,d\Omega} &=& 
R_{np}^N\frac{1}{N!Z!A}\left(\frac{4\pi P_0^3}{3[2m^3(E-E_C)]
^{1/2}}\right)^{A-1}\nonumber\\
&&\times\left(\frac{d^2N(1,0,E)}{dE\,d\Omega}\right)^A,
\label{eqnCoal}
\end{eqnarray}
where the double differential multiplicity for a cluster of mass number $A$ 
containing $Z$ protons and $N$ neutrons and having a Coulomb-corrected energy 
$E_A$, is related to the proton double differential multiplicity at the same 
Coulomb corrected energy per nucleon,  $E-E_C$, where $E_C$ is the Coulomb 
barrier for proton emission.  $R_{np}$ is the neutron to proton ratio.

\begin{figure}
\epsfig{file=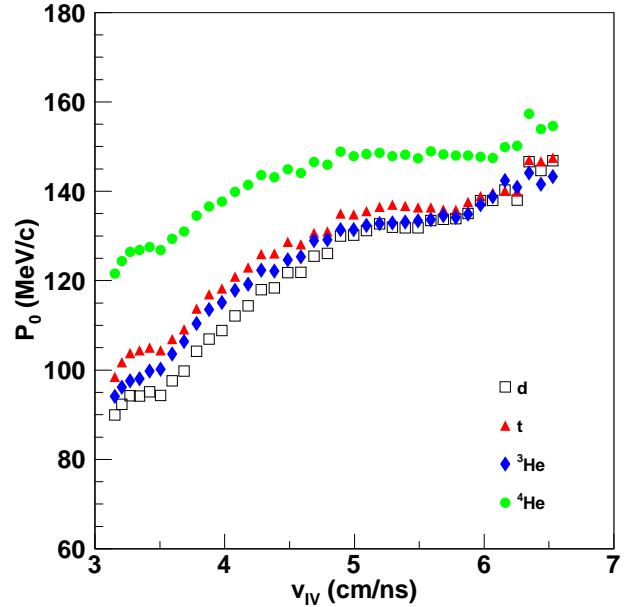,width=9.2cm,angle=0}
\caption{(Color online) Coalescence parameters, $P_0$ as a function of surface 
velocity in the intermediate velocity source frame. Reaction: 47 $A$ MeV $^{40}$Ar
+ $^{112}$Sn}
\label{fig1}
\end{figure}

Application of the coalescence model requires knowledge of cluster, 
neutron and proton differential cross sections with proper absolute 
normalizations. In this work absolute measured multiplicities for the 
selected violent events are employed. The neutron spectra are not measured. 
However, since within the framework of the coalescence model the yield 
ratios of two isotopes that differ by one neutron are determined by their 
binding energies and the $n/p$ ratio in the coalescence volume, we have 
used the observed triton to $^3$He yield ratio to derive the $n/p$
ratio used in this analysis. Because our goal was to derive information 
on the density and temperature evolution of the emitting system, our 
analysis was not limited to determining an average $P_0$ value.  
Instead, as in our previous 
studies \cite{kowalski07,natowitz10,roepke09,qin08,hagel00}, results for 
$d,\, t, \,^3$He, and $^4$He were derived as a function of the surface velocity 
in the intermediate velocity source frame, i.e.,  the velocity of the 
emerging particle at the nuclear surface, prior to Coulomb acceleration. 
Results for the reaction of $^{40}$Ar + $^{112}$Sn are presented in Fig.~\ref{fig1}. 
Values of $P_0$ for the other reactions are very similar. The low velocity 
cut off is imposed to avoid significant contamination from particles 
evaporated from the TLF source. The velocity of 6.6 cm/ns in the IV frame 
corresponds to the beam velocity and the rise in $P_0$ seen in the Figure 
indicates that these high velocity particles may not be amenable to an 
equilibrium treatment.

\begin{figure}
\epsfig{file=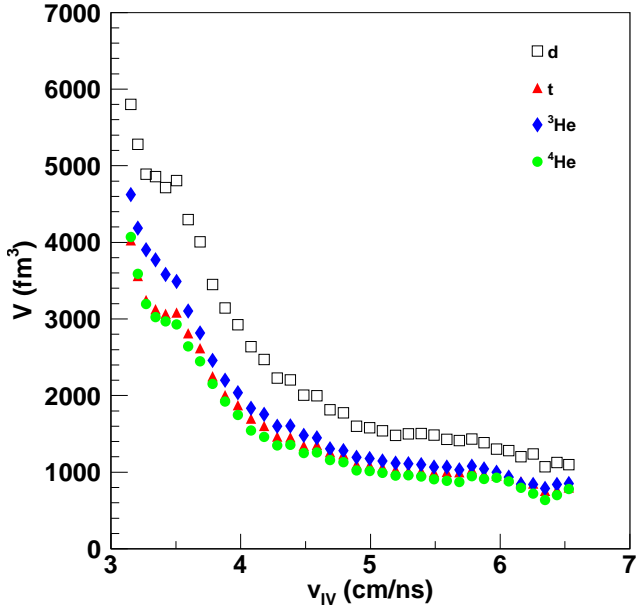,width=9.2cm,angle=0}
\caption{(Color online) Coalescence model volumes as a function of velocity 
in the intermediate velocity source frame.  Reaction: 47 $A$ MeV $^{40}$Ar
+ $^{112}$Sn}
\label{fig2}
\end{figure}

In the Mekjian model thermal and chemical equilibrium determines coalescence 
yields of all species. Under these assumptions there is a direct relationship 
between the derived radius in momentum space and the volume of the emitting 
system. In terms of the $P_0$ derived from Eq.~(\ref{eqnCoal}) and assuming 
a spherical source,
\begin{equation}
\label{vol}
V=\frac{3h^3}{4\pi P_0^3}\,\left[ \frac{Z!N!A^3}{2^A} (2s+1)e^{\frac{E_0}{T}} \right]^{\frac{1}{A-1}}
\,,
\end{equation}
where $h$ is Plancks constant, and $Z$, $N$, and $A$ are the same as in 
Eq.~(\ref{eqnCoal}), $E_0$ is the ground state binding energy, $s$ the spin 
of the emitted cluster, and $T$ is the temperature. Thus the volume can be 
derived from the observed $P_0$ and temperature values assuming a spherical 
source.  We note that this volume is a free volume.  From the relevant $P_0$ 
values we then determined volumes using Eq.~(\ref{vol}).  See Fig.~\ref{fig2} 
where, again,  results for the reaction $^{40}$Ar + $^{112}$Sn are presented.

A comparison of these volumes indicates good agreement for $t$, $^3$He 
and $^4$He. Volumes derived from the deuteron data are typically somewhat 
larger. This larger apparent volume for deuterons appears to reflect the 
fragility of the deuteron and its survival probability once 
formed \cite{cervesato92}. For this reason we have used average volumes 
derived from the $A\geq 3$ clusters to calculate the densities. Given that 
mass is removed from the system during the evolution, we determined the 
relevant masses for each volume by assuming that the initial mass of the 
source was that obtained from the source fitting analysis and then 
deriving the mass remaining at a given $v_{\rm surf}$ from the observed 
energy spectra. This is also an averaging process and ignores fluctuations.
Once these masses were known they were used to determine an excluded volume 
for the particles.  Addition of this excluded volume to the free volume 
produced the total volumes needed for the density calculations.  These were  
determined by dividing the remaining masses  by the total volume. This 
was done as a function of $v_{\rm surf}$.  This excluded volume correction is 
small for  low  $v_{\rm surf}$ but increases with increasing $v_{\rm surf}$.  
Densities are those of total number of nucleons, including the nucleons 
bound in clusters, per fm$^3$.

\section{Results}
\subsection{Temperatures and Densities}
Inspection of the results for the four different systems studied revealed 
that, as a function of surface velocity,  the temperatures, densities and 
equilibrium constants  for all systems are the same within statistical 
uncertainties. This indicates that the time evolution of the systems 
studied is very similar \cite{hagel00,wang25}. Therefore we have combined 
values from all systems to determine the temperatures and densities  
reported in this paper.   

\begin{figure}
\epsfig{file=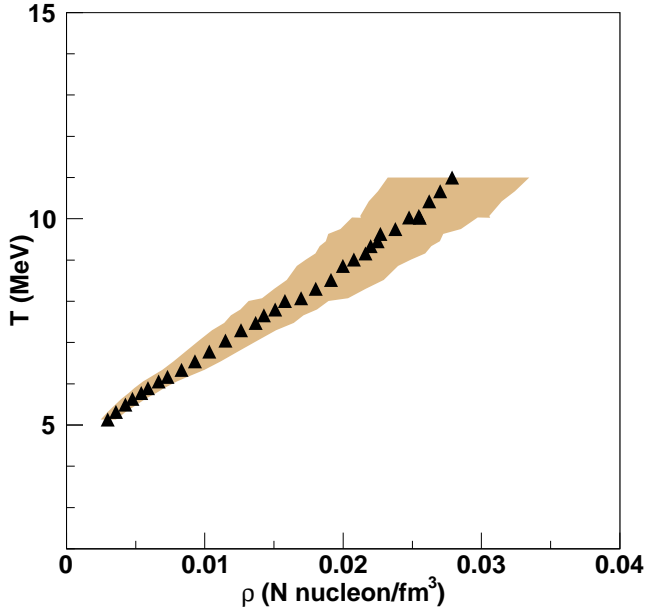,width=9.2cm,angle=0}
\caption{(Color online) Temperatures and densities sampled by cluster 
emission from the expanding IV source.}
\label{fig3}
\end{figure}

We present in Fig.~\ref{fig3} the experimentally derived density and 
temperature evolution for the IV source.  Estimated errors on the 
temperatures are 10\% at low density evolving to 15\% at the higher densities.
In deriving our results we have invoked the correlation between average 
surface velocity and emission time for the early emitted particles on which 
we focus our attention.  This is seen clearly in transport model 
calculations~\cite{wang25}.  It is certainly true that density and 
temperatures corresponding to the different surface velocity bins are 
weighted averages over the underlying distributions.  There will be 
fluctuations in $T$ and $\rho$ present at the time of particle emission. 
We assume these to be symmetric about the most probable value.  There will 
then be additional fluctuations induced by mixing.  Using the AMD model of 
Ono, and assuming at each time a single (time-decreasing) temperature we 
estimate that, for the higher surface velocities, the weighted average 
temperatures are $\leq 10\%$ lower than the input temperatures.  This effect 
decreases at lower surface velocities where the rate of decrease of 
temperature with velocity is less.  These estimates are included in our 
estimates of the temperature uncertainty.  The error in the derivation of 
the density arises from the uncertainty on the volume which is dominated by 
the uncertainty in temperature and the uncertainty in source mass derived 
from source fitting to complex spectra.  The estimated errors on the 
densities are $\pm 17\%$.  

\subsection{Isoscaling}
Considering only bulk properties of homogeneous nuclear matter, and the 
standard thermodynamic relations that relate the Free energy, the Internal 
energy, and the entropy, as a function of $\rho, \delta, T$ we can expand 
with respect to $\delta$ to define the corresponding expressions for 
$F_{\rm sym}(\rho,T)$ and
$S_{\rm sym}(\rho,T)=-\partial F_{\rm sym}(\rho,T)/\partial T$. This is 
analogous to the expressions given above for $E_{\rm sym}(\rho,T)$  

The isoscaling technique has been used to derive information on $F_{\rm sym}$, 
the symmetry Free energy of the system.  The isoscaling technique consists of 
measuring cluster yield ratios for two different  excited systems 
(denoted by the index 1 and 2), having the same temperature (and density) 
and  the same atomic number but differing in the $Z/A$ ratio.  For such 
systems liquid drop model binding energy terms other than those related 
to the symmetry energy coefficient essentially cancel and one can 
write \cite{tsang01,souza08}.  
\begin{eqnarray}
\label{eq4}
\frac{Y_2}{Y_1}&=&
Ce^{((\mu_2(n)-\mu_1(n))N+(\mu_2(p)-\mu_1(p))Z)/T} \nonumber \\
&=&Ce^{\alpha N+\beta Z}
\end{eqnarray}
where $C$ is a constant and $\mu(n)$ and $\mu(p)$ are the neutron and 
proton chemical potentials.  The isoscaling parameters 
$\alpha=\big(\mu_2(n) - \mu_1(n)\big)/T$ and $\beta= \big(\mu_2(p) - \mu_1(p)\big)/T$ , 
representing the difference in chemical potential between the two systems, 
may be extracted from suitable plots of yield ratios. Either parameter may then 
be related to the symmetry Free energy, $F_{\rm sym}$. We take the $\alpha$ 
parameter, which is expected to be less sensitive to residual Coulomb effects. 
Addressing specifically the symmetry Free energy and adopting the usual 
convention that system 2 is richer in neutrons than system 1,
one can write
\begin{equation}
\alpha=4F_{\rm sym}\left[\left(\frac{Z_1}{A_1}\right)^2-\left(\frac{Z_2}{A_2}\right)^2\right]/T\,,
\end{equation}
where $Z$ is the atomic number and $A$ is the mass number of the emitter 
\cite{natowitz10,tsang01,souza08}. Thus, $F_{\rm sym}$ may be derived directly 
from determinations of system temperatures, $Z/A$ ratios, and 
isoscaling parameters. We emphasize that the present analysis is carried 
out for light species characteristic of the nuclear gas rather than, as 
in most previous analyses, for the intermediate mass fragments thought 
to be characteristic of the nuclear liquid. In this work we employ Eq.~(\ref{eq4})
with experimentally determined isoscaling parameters, $\alpha$,  
temperatures, $T$, and $Z/A$ ratios to determine the symmetry 
Free energy coefficient, $F_{\rm sym}$.  

\begin{figure}
\epsfig{file=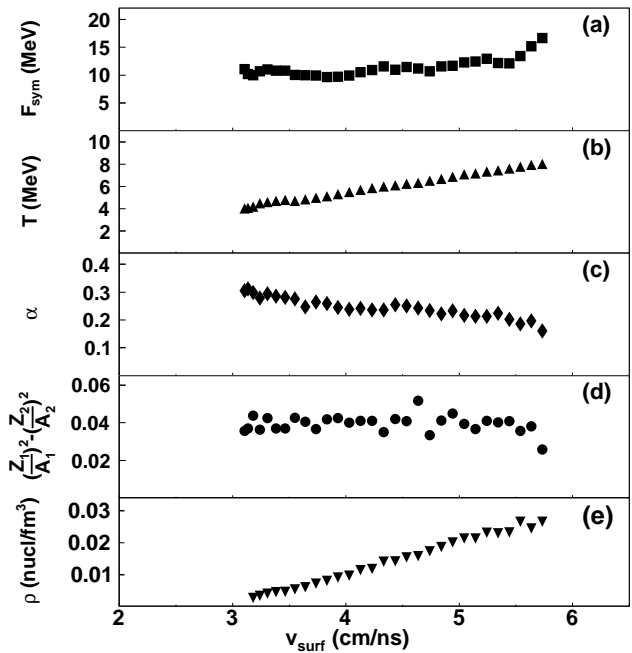,width=9.2cm,angle=0}
\caption{Parameters of the analysis.  (a) Symmetry Free energy coefficient, (b) Temperature, (c) Isoscaling parameter, $\alpha$, 
(d) $(\frac{Z_1}{A_1})^2 - (\frac{Z_2}{A_2})^2 $, (e) Density
}
\label{fig4}
\end{figure}

Fig.~\ref{fig4} (a-e) presents the results of our isoscaling analysis. The 
experimentally derived evolution of density, temperature, 
$[(Z_{1}/A_{1})^2- (Z_{2}/A_{2})^2]$, the isoscaling 
parameter $\alpha$, and the Free symmetry energy coefficient are 
presented as a function of surface velocity in the IV frame. The correlations 
between these parameters become apparent in this figure.  For the 
determination of the Free symmetry energies we have restricted ourselves 
to isoscaling results leading to alpha parameters with less than 15 \% 
uncertainties.  This restricts the temperature and density range for which 
the Free symmetry energies are reported in figure 4. 

We see that $T$ and $\rho$ increase with increasing $v_{\rm surf}$. At the same 
time there is a decline in the isoscaling coefficient $\alpha$.  Combined, 
the terms of Eq.~(\ref{eq4}) lead to values of $F_{\rm sym}$ that rise 
gently with $v_{\rm surf}$. 

\subsection{Free Symmetry Energy}
In Figure 5 the Free symmetry energy coefficients determined from the 
isoscaling analysis are plotted. Also shown are calculated Free symmetry 
energy coefficients obtained from a quantum statistical approach that 
includes medium modifications of the cluster binding 
energies \cite{roepke09,roepke11,typel10,sumiyoshi95,ref13}. A few-body 
Schr\"odinger equation has been solved that contains single-particle 
self-energy shifts as well as Pauli blocking factors. For given $T$ and 
$\mu_p, \mu_n$, the nucleon density is calculated. To obtain a thermodynamic 
potential, the Free energy as function of $T, n_n, n_p$ is evaluated by 
integration.  The further thermodynamic quantities are determined from the 
Free energy in a consistent way.

\begin{figure}
\epsfig{file=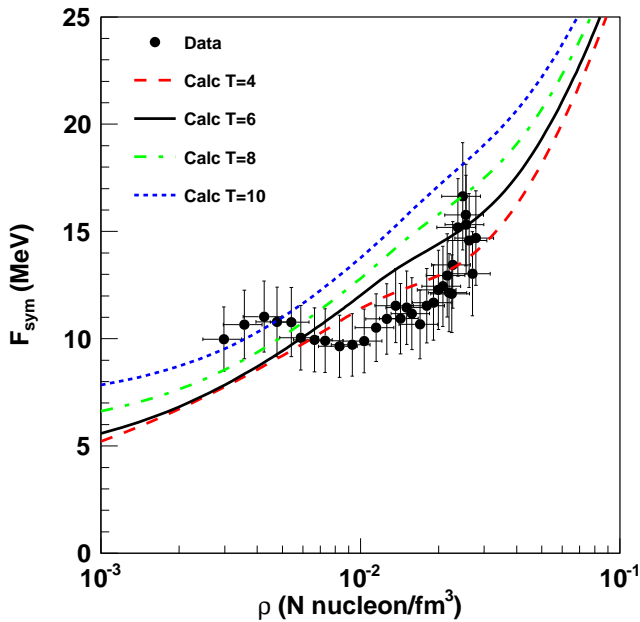,width=9.2cm,angle=0}
\caption{(Color online) Free symmetry energy $vs.$ density.  Symbols show 
experimental points.  Lines are calculated values for $T$ =4, 6, 8 and 
10 MeV~\cite{roepke09,roepke11,typel10,sumiyoshi95,ref13}}
\label{fig5}
\end{figure}

Isotherms of the Free symmetry energy at $T$ = 4, 6, 8 and 10 MeV are also 
shown in Fig.~5.  Note that the experimental data correspond to changing 
temperatures as seen in Fig. 4.  Within the error bars, the general behavior 
of the experimental data is fairly well reproduced by the calculations. The 
data are somewhat below the calculated values, in particular in the 
intermediate region.

\begin{figure}
\epsfig{file=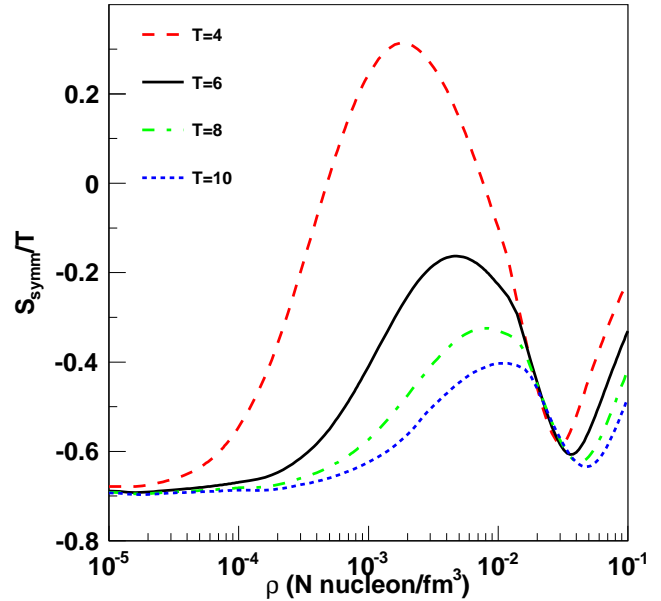,width=9.2cm,angle=0}
\caption{(Color online) Symmetry entropy vs density.  Symbols show 
experimental points and lines show calculations for T=4, 6, 8 and 10 MeV.}
\label{fig6}
\end{figure}

\begin{figure}
\epsfig{file=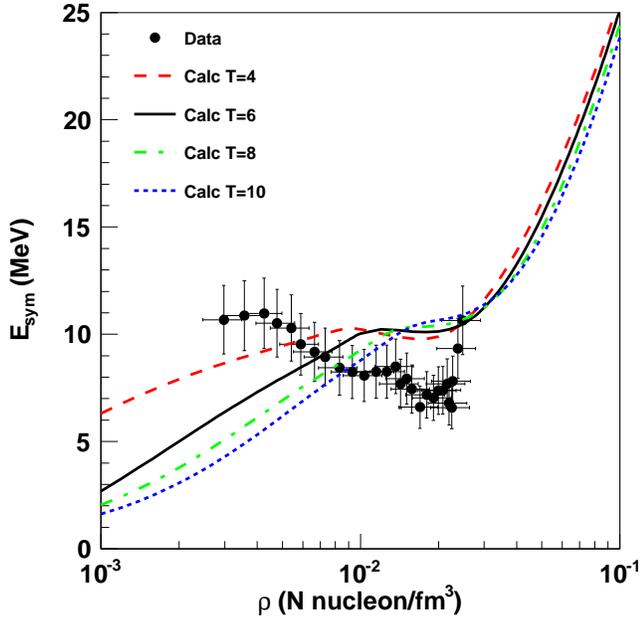,width=9.2cm,angle=0}
\caption{(Color online) Symmetry energy vs density.  Symbols show 
experimental points and lines show calculations for T=4, 6, 8 and 10 MeV.}
\label{fig7}
\end{figure}

To derive the values of the Internal symmetry energy from the Free symmetry 
energy, the symmetry entropy must be known.  The calculations of the symmetry 
entropies have been performed using the model of 
Typel~\textit{et al.}~\cite{typel10} where the Free energy, the 
Internal energy and the entropy have been given for symmetric matter.  
For pure neutron matter, the RMF approach can be used because there is no 
cluster formation.  

We calculated the entropy within the QS approach by differentiating the Free 
energy with respect to the temperature.  In Figure 6 we present 
the symmetry entropy, i.e. the difference between the entroy of neutron 
matter and that of symmetric matter, for different temperatures using the 
recent quantum statistical calculations of momentum dependent energy shifts 
of light clusters~\cite{roepke11}.

In contrast to the mixing entropy that leads to a larger entropy for 
uncorrelated symmetric matter in comparison with pure neutron matter, the 
formation of correlations, in particular clusters, will reduce the entropy 
in symmetric matter, see also Fig.~9 of Ref.~\cite{typel10}.  For 
parameter values for which the yields of free nucleons in symmetric matter 
are small, the symmetry entropy may become positive.  The fraction of 
nucleons bound in clusters can decrease due to increasing temperature 
or the dissolution of bound states at high densities due to the Pauli 
blocking.  Then, the symmetric matter recovers its larger entropy so that 
the symmetry entropy becomes negative.  

The corresponding 
results for the symmetry energy are shown in Fig. 7.  Experimental as well as 
calculated values indicate a large value of the symmetry energy also in the 
low-density region since the alpha-fraction in symmetric matter becomes 
large at low temperatures, and the value of the binding energy per nucleon 
determines the Internal energy.  This effect in not included in quasiparticle 
approaches such as standard relativistic mean field calculations because 
few-body correlations are neglected. 

\section{Summary and Conclusions}
The NIMROD multi-detector at Texas A\&M University has been employed to 
extend our earlier measurements of symmetry energy coefficients in low 
density clustered nuclear matter. Yields of light particles produced in 
the collisions of 47 $A$ MeV  $^{ 40}$Ar with $^{112}$Sn ,$^{124}$Sn  
and  $^{64}$Zn with $^{112}$Sn, $^{124}$Sn were employed in Thermal 
coalescence model analyses to derive densities and temperatures of the 
evolving emitting systems. Isoscaling analyses were used to determine 
the Free symmetry energies of these systems.  Comparisons of the 
experimental values are made with those of calculations made using a model 
which incorporates medium modifications of cluster binding energies.  
The model calculated symmetry entropies have been used together with the 
experimental Free symmetry energies to derive symmetry energies of nuclear 
matter at densities of $0.03 \leq \rho/\rho_0 \leq 0.2$  and temperatures in 
the range 5 to 10 MeV.
 
\section{Acknowledgements}
This work was supported by the United States Department of Energy 
under Grant \# DE-FG03- 93ER40773 and by The Robert A. Welch 
Foundation under Grant \# A0330.

\end{document}